\documentstyle[art8,aaspp4,flushrt,tighten]{article}

\newcommand{\lea}{\lesssim}
\newcommand{\gap}{\gtrsim}

\newcommand{\vth}{v_{\rm th}}

\newcommand{\rwd}{\, r_{\rm wd}}
\newcommand{\rwdt}{r_{\rm wd}}
\newcommand{\Mwd}{M_{\rm wd}}
\newcommand{\Msun}{\,{\rm M}_\odot}
\newcommand{\Lsun}{\,{\rm L}_\odot}
\newcommand{\msy}{\, {\rm M}_\odot \, {\rm yr}^{-1}}
\newcommand{\rd}{r_{\rm d}}

\newcommand{\Gammad}{{\Gamma}_{\rm d}}
\newcommand{\cri}{_{\rm c}}
\newcommand{\xx}{x}
\newcommand{\XX}{X}
\newcommand{\FF}{{\tilde F}}
\newcommand{\Flux}{F}
\newcommand{\Ld}{L_{\rm d}}
\newcommand{\sigmae}{\sigma_{\rm e}}

\received{}
\accepted{}
\journalid{}{}
\articleid{}{}

\slugcomment{Submitted to the Astrophysical Journal}

\begin{document}

\title{Dynamics of Line-Driven Winds from Disks in Cataclysmic
Variables\\ II. Mass Loss Rates and Velocity Laws}

\author{Achim Feldmeier}

\affil{Department of Physics \& Astronomy, University of Kentucky,
Lexington, KY 40506-0055, USA, E-mail: {\tt achim@pa.uky.edu}}

\and

\author{Isaac Shlosman}

\affil{Department of Physics \& Astronomy, University of Kentucky,
Lexington, KY 40506-0055, USA, E-mail: {\tt shlosman@pa.uky.edu}}

\and

\author{Peter Vitello}

\affil{Lawrence Livermore National Laboratory, L-14, P. O. Box 808,
Livermore, CA 94550, USA, E-mail: {\tt vitello@spaniel.llnl.gov}}

\begin{abstract}

We analyze the dynamics of 2-D stationary, line-driven winds from
accretion disks in cataclysmic variable (CV) stars, by generalizing
the Castor, Abbott and Klein (CAK) theory. In the first paper, we have
solved the wind Euler equation, derived its two eigenvalues, and
addressed the solution topology and wind geometry. Here, we focus on
mass loss rates and velocity laws of the wind.

We find that disk winds, even in luminous novalike variables, have low
optical depth, even in the strongest driving resonance lines. This
suggests that thick-to-thin transitions in these lines occur in the
wind. For disks with a realistic radial temperature law, the mass loss
is dominated by gas emanating from the inner decade in radius. The
total mass loss rate associated with the wind from a disk of
luminosity $10 \Lsun$ is $\sim 10^{-12} \msy$, or $10^{-4}$ of the
mass accretion rate. This is one order of magnitude below the lower
limit obtained from fitting P~Cygni line profiles using kinematical
wind models, when the ionizing flux shortwards of the Lyman edge is
supressed. The difficulties associated with such small mass loss rates
for line-driven winds from disks in CVs are principal, and confirm our
previous work on this subject. We conjecture that this issue may be
resolved by detailed non-LTE calculations of the CAK line force within 
the context of CV disk winds, and/or better accounting for the disk 
energy distribution and wind ionization structure.

We find that the wind velocity profile is well approximated by the
empirical law used in kinematical modeling. The acceleration
length scale is given by the footpoint radius of the wind streamline
in the disk. This suggests an upper limit of $\sim 10 \rwd$ to the
acceleration scale, which is smaller by factors of a few as compared
to values derived from line fitting.

\end{abstract}


\section{Introduction}

Line-driven winds (hereafter LDWs) are expected around luminous
objects whose spectra peak in the UV, such as OB~stars and accretion
disks, stellar and galactic. Feldmeier \& Shlosman (1999, Paper~I)
have investigated a 2-D analytical model of LDWs from disks in
cataclysmic variables (CVs) characterized by a large mass transfer
rate from the secondary to the white dwarf. Such CVs, i.e., novalike
variables and dwarf novae in outburst, show clear signs of outflows
driven by radiation pressure (Paper~I and refs.~therein).

Recent numerical simulations of time-dependent 2-D disk winds by
Proga, Stone, \& Drew (1998; PSD hereafter) largely confirmed the
previous kinematical studies by Shlosman \& Vitello (1993) and Knigge,
Woods, \& Drew (1995), and delineated a number of empirical
relationships which require further physical explanation. Paper~I has
addressed the 2-D geometry of the wind streamlines and the topology of
solutions to the wind momentum equation. In particular, a comparison
with the PSD model was made as well as a comparison with the 1-D LDWs
from OB~stars.

The main results of Paper~I are as follows. Firstly, the solutions to
the wind momentum equation are characterized by two eigenvalues, the
mass loss rate and the flow tilt angle with the disk, in the presence
of a realistic radiation field above the disk. The existence of the
second eigenvalue is a reflection of the multi-dimensional nature of a
disk wind. The wind itself appears to be collimated to a certain
degree, i.e., the wind collimation angle with the rotation axis
(semi-opening angle) is about $10\arcdeg$ for the wind launched within
4 white dwarf radii, and about $30\arcdeg$ for the outer wind, for the
Shakura-Sunyaev (1973; SHS hereafter) disk. Furthermore, the wind
collimation depends solely on the radial temperature stratification in
the disk, unless there is an additional degree of freedom such as a
central luminosity associated with nuclear burning on the surface of
the white dwarf. The above degree of collimation for disk winds in CVs
should be taken with caution at large distances from the disk.

Secondly, a major distinction between stellar and disk LDWs is the
existence of maxima in both the gravity and the radiation flux above
the disk. This behavior of gravity and radiation flux results in
profound topological differences in the solutions to the stellar and
disk wind momentum equations. We find that two distinct regions of
disk wind exist, the inner and outer winds. The critical point of the
outer wind lies close to the disk photosphere and to the sonic point,
upstream of the top of the gravity `hill'. This proximity of the
critical and sonic points is typical of LDWs from O~stars as well. On
the other hand, for the inner wind, the critical point lies far away
from the sonic point, beyond the gravity hill.

Observationally, the mass loss rates from CV disk winds are poorly
constrained. This is mainly a consequence of uncertainties in the
ionizing fluxes from different components in the system, and,
therefore, of the ionization stratification of the wind. Neglecting
the boundary layer and assuming a local blackbody radiation from the
disk, Vitello \& Shlosman (1993) and Knigge~et~al.~(1995) find wind
mass loss rates $\sim 1$\% of the accretion rate, by fitting observed
P~Cygni line profiles. For a system of luminosity $10\Lsun$, this
corresponds to a mass loss rate of about $2\times 10^{-10} \msy$.

These mass loss rates are {\it upper} limits for the following
reason. The continuum radiation field shortwards of the Lyman edge is
often found (Polidan, Mauche, \& Wade 1990; Long~et~al.~1991, 1994;
van Teeseling, Verbunt, \& Heise 1993; Knigge~et~al.~1997) to be
highly supressed compared to blackbody emission or stellar
photospheric fluxes, or is even absent. This drastic reduction in
ionizing flux allows to reduce the electron density, and, therefore,
$\dot M$, while maintaining the same degree of ionization in the wind.

On the other hand, a reasonable {\it lower} limit to the wind mass
loss rate from luminous CV disks was found by Prinja \& Rosen (1995).
They argued that the product $\dot M q$, where $q$ is the ionization
fraction of C~{\sc iv}, lies between $5\times 10^{-13}$ and $1\times
10^{-11} \msy$ for ten dwarf novae and novalike variables with
high-resolution IUE spectra (see also Mauche \& Raymond 1987; Hoare \&
Drew 1993). This results in a lower limit of $\dot M\sim 5\times
10^{-13} \msy$. Note that this value is still model dependent to some
degree, since Prinja \& Rosen assumed a constant ionization fraction
throughout the wind, and a purely radial flow.

In this paper, we focus on the mass loss rate and the velocity law of
disk LDWs in CVs. We employ the main assumption of Paper~I that the
wind streamlines are contained in straight cones, whose collimation
angles are eigenvalues of the flow. This approximation is acceptable
close to the disk (e.g., as shown by PSD) and allows us to make a
meaningful determination of the mass loss rate (which is constrained
close to the disk photosphere as well), and of the initial velocity
law. A similar conclusion was reached from line fitting using 2-D wind
kinematics (Vitello \& Shlosman 1993; Knigge~et~al.~1995). On the
other hand, we expect both the centrifugal forces and the polar
component of the line force to bend the streamlines at large distances
from the disk, and to influence the terminal velocity of the flow.

This paper is organized as follows. Section~2 reviews the mass loss
rates derived from CAK theory, Section~3 shows that CV disk winds are
expected to have low optical depth even in strong lines. Sections~4
and 5 derive the mass loss rates and velocity laws for the disk wind
model. Sections~6 and 7 discuss and summarize our results.

\section{CAK mass loss rates for stars and CV disks}

The CAK line force for stellar winds is fully determined by two
parameters, the power exponent $\alpha$ and the mass absorption
coefficient of the strongest driving line, $\kappa_0$ (Paper~I; Puls,
Springmann, \& Lennon 1999).  Instead of using $\kappa_0$ directly,
CAK parameterize the line force per unit mass as $g_{\rm L}= M(t) \,
g_{\rm e}$, where $M(t)\equiv k\,t^{-\alpha}$ is the so-called force
multiplier, and $g_{\rm e}$ is the force due to electron
scattering. The optical depth $t$ refers to a line with
$\kappa=\sigmae$, where $\sigmae$ is the electron scattering
coefficient. The parameter $k$ is given in terms of $\kappa_0$ by
 \begin{equation}
 \label{kofkappa0}
 k={\Gamma(\alpha) \over 1-\alpha} \,{\vth \over c}\, \left({\kappa_0
\over \sigmae}\right)^{1-\alpha},
 \end{equation} 
 \noindent where $\Gamma(\alpha)$ is the complete Gamma function, and
$\vth$ is the thermal speed of carbon ions (CAK) or hydrogen (Abbott
1982). The above parameterization for $M(t)$ as a power law, however,
does not account correctly for optically thin winds, where the force
multiplier saturates at some value $M_{\rm max}(t) \equiv Q\sim 2,000$
for a gas of solar composition (Abbott 1982).  This particular value
of $Q$ was determined for O~stars. However, already for winds from
B~stars near-the-main sequence, modeling of X-ray spectra suggests
that mass loss rates inferred from the standard theory can be
substantially underestimated (Cassinelli 1994;
Cassinelli~et~al.~1994). In terms of $Q$, $\kappa_0$ is given by
(Gayley 1995),
 \begin{equation}
 \label{kappa0ofq}
 {\kappa_0\,\vth \over \sigmae \,c}=\Gamma(\alpha)^{-{1 \over
1-\alpha}}\, Q.
 \end{equation} 
 \noindent Inserting the eigenvalue $E\cri$ from eq.~(I.8b; hereafter
tables, figures and equations from Paper~I have a roman prefix `I')
into (I.6), the CAK mass loss rate from O~star winds takes the compact
form,
 \begin{equation}
 \label{gayleysmasslosseq}
 {\dot M} = {\alpha \over 1-\alpha} \, \left({Q\,\Gamma \over
1-\Gamma}\right)^{1-\alpha \over \alpha} \, {L \over c^2}.
  \end{equation}
 \noindent This expression is valid as long as
$Q\,\Gamma/(1-\Gamma)>1$.  Otherwise, the gravity prevails and no wind
solution exists.  For the above value of $Q$, and assuming
$\alpha=2/3$ and an Eddington factor $\Gamma=0.5$ of an O supergiants,
eq.~(\ref{gayleysmasslosseq}) gives $\dot M \sim 90 \, L/c^2$. The
mass loss rate from CAK theory agrees thus well with the estimate from
the single scattering limit, $\dot M = (c/v_\infty)\, (L/c^2) \sim 100
\, L/c^2$, for typical terminal speeds $v_\infty\sim 3,000~{\rm km \,
s^{-1}}$. However, agreement between both mass loss rates is solely
due to the fact that $\Gamma$ is close to unity.

If, alternatively, $\Gamma \ll 1$, the CAK mass loss rate falls well
below the single scattering limit. As we show in the next section, the
mass loss of a disk LDW is given again by
eq.~(\ref{gayleysmasslosseq}), up to correction factors of order
unity. Even for the brightest CVs, i.e., novalike variables and dwarf
novae in eruption, which experience LDWs, $Q \,\Gamma\sim 1$ (applying
the O~star value of $Q$). Hence, $\dot M \sim L/c^2$ from
eq.~(\ref{gayleysmasslosseq}), whereas the single scattering limit
gives $\dot M \sim 60\, L/c^2$, for $v_\infty \sim 5,000~{\rm km\,
s^{-1}}$. Since in thin LDWs the probability for a photon to be
scattered by a line is less than unity, the estimate from the single
scattering is way too high (see also Puls, Springmann, \& Owocki
1998).

\section{Disk wind optical depths}

A number of fundamental differences exist between stellar and disk
LDWs in CVs, some of them discussed in Paper~I. Here we show that
optical depths for CV winds are more typical of thin winds, e.g., of
B~stars near-the-main sequence, than of more extensive supergiant
winds.

\subsection{Disk wind geometry and radiation field}

The disk wind geometry is described in Paper~I, and we repeat here
only the essential assumptions and make necessary definitions.  A flow
streamline is a helix which is contained within a {\it straight} cone
(Fig.~1). The footpoint radius of a streamline in the disk is $r_0$,
the tilt of the cone with the disk is $\lambda$, and $\xx$ is the
distance along the cone.  We neglect pressure forces, and assume that
the azimuthal velocity is determined by angular momentum conservation
above the disk, and by Keplerian rotation within the disk. The only
remaining velocity component is $v_\xx$, which points upwards along
the cone. We introduce a normalized coordinate $\XX\equiv \xx/r_0$.
The velocity $V$ is normalized to the local escape speed, and the flow
acceleration becomes $W'=2V\,{dV/d\XX}$. Lastly, we introduce the
radiation flux ${\bf\Flux}$ above the disk and the flux ${\bf\FF}$
normalized to the footpoint flux of the streamline, as well as their
projected counterparts $\Flux_\xx$ and $\FF_\xx$ along the wind cone
(Section 3.2 of Paper~I).

\subsection{Semi-transparent winds from disks in CVs?}
\label{whythin}

In this section, we make use of a simplified Euler equation for the
disk LDW to show explicitly that low $\Gamma$ factors in CV disks
imply low optical depths in the wind. By doing so, we neglect factors
of order unity from the angle integration in eq.~(I.2), in replacing
$\tau_\gamma$ by $\tau_\xx$. For simplicity, only the disk with $\Flux
\propto r_0^{-2}$ (hereafter `Newtonian') is considered. The line
force is then
 \begin{equation}
 g_{\rm L} = {\sigmae \Flux_\xx(r_0,\XX) \over c} \, M(t_\xx) =
{\sigmae \Flux_z(r_0,0) \over c} \, \FF_\xx(r_0,\XX) \, M(t_\xx),
 \end{equation}
 \noindent The Euler equation for the disk wind, in the limit of zero
sound speed, and neglecting the force due to electron scattering
because of small $\Gamma$ above the disk, becomes (with $\Mwd$ being
the white dwarf mass),
 \begin{equation}
 \label{euler_justsowind}
 v_\xx v'_\xx= - {G\,\Mwd \over r_0^2} \left[g(\XX)- \Gamma(r_0)
\FF_\xx(r_0,\XX) M(t_\xx)\right],
 \end{equation}
 \noindent where the effective gravity $g$ was defined in (I.9), and
we introduced
 \begin{equation}
 \label{localeddington}
 \Gamma(r_0)\equiv {\sigmae \,\Flux_z(r_0,0)\, r_0^2 \over
c\,G\,\Mwd}.  \end{equation}
 \noindent For the Newtonian disk, $\Gamma= \sigmae \, \Ld / [4\pi\,
c\,G\,\Mwd \ln(\rd/\rwd)]$ becomes independent of $r_0$. Here, $\Ld$
is the disk luminosity and $\rd$ and $\rwd$ are the outer and inner
disk radii, $\rwd$ being the white dwarf radius.  Using typical
parameters for novalike CVs, $\Ld=10 \, \Lsun$, $\Mwd=0.7 \, \Msun$,
and $\rd/\rwd=30$, one has that $\Gamma \simeq 10^{-4}$.

For a stationary wind solution to exist, the right-hand-side of
eq.~(\ref{euler_justsowind}) has to be positive. This poses a
constraint on $M(t)$ and therefore on $t$. Namely, the maximum of
reduced gravity $g$ lies between $2/(3\sqrt{3}) \simeq 0.38$ for
$\lambda=90\arcdeg$ and $4/27 \simeq 0.15$ for $\lambda=0\arcdeg$.
Since $\FF_\xx$ is of order unity (see Fig.~I.5), $M(t)$ approaches
its maximum value, $Q$, in regions of large gravity, and stays
constant thereafter. In other words, because $\Gamma$ is so small for
CV disks, the wind solution barely `makes it' over the gravity hill.

This saturation effect in $M(t)$ happens when the strongest driving
line in the wind becomes optically thin, at about $t\sim 10^{-7}$. If
this thick-to-thin transition occurs before or at the critical point
of the flow, the wind solution is lost. This possiblity cannot be
excluded in our model due to the rapid change of the velocity gradient
(and hence of $t$) in the vicinity of the critical point. The
consequences of this effect on the feasibility of LDWs are discussed
in Section~\ref{discussion}.

The situation is fundamentally different for dense winds of O~stars,
where the Euler equation reads
 \begin{equation}
 vv'= - {G\,\Mwd \over r^2} \left(1-\Gamma [M(t)+1]\right).
 \end{equation}
 \noindent Assuming $\Gamma > 0.1$ for O~stars, there is a wide range
in $M(t)$ for a stationary solution to exist, namely from $\sim
10-2,000$. The highest mass loss rate solution (hence, the slowest
wind) is characterized by the lowest allowable $M(t)\sim 10$. The
permitted range in $M(t)$ corresponds to an even wider range in $t$
(because $\alpha < 1$), $t\sim 10^{-7}-10^{-3}$. We further quantify
these arguments in Appendix A.

Our estimate for $t$ in CV winds contradicts the claim by Murray \&
Chang (1996) that the optical depth parameter $t$ is similar for CV
disk and O~star winds.

\section{Disk wind mass loss rates}
\label{diskwindmasslossrates}

\subsection{Vertical disk wind}

The mass loss rate of the vertical wind above an isothermal disk is
determined by the eigenvalue $E\cri$ in eq.~(I.18b). From eq.~(I.17)
one finds,
 \begin{equation}
 \label{masslossfromverticalwind1}
 \dot M= {\alpha\over 1-\alpha} \left({3\sqrt{3} \,c\,\sigmae\, Q\over
8\pi \, G\,\Mwd}\right)^{1-\alpha\over\alpha} \, D \, \left({\Ld \over
c^2}\right)^{1\over \alpha}.
 \end{equation}
 \noindent The dimensionless constant $D$ is given by
 \begin{equation}
 \label{diskwindareafactor}
 D=\int\limits_{r_1}^{r_2} {dr \over r} \left({4\pi \, r^2 \,
\Flux_z(r,0) \over \Ld}\right)^{1 \over \alpha},
 \end{equation}
 \noindent where $r_1$ and $r_2$ are the inner and outer radii of the
wind base in the disk, respectively. Making the plausible assumptions
that $\rd \gg \rwd$, $r_2 \gg r_1$, and $r_2\simeq \rd$ (both of the
latter radii are determined essentially by the temperature dropping
below $10^4$~K), we estimate $D\simeq 1$ for $\alpha=1/2$ and
2/3. Introducing the disk Eddington factor
 \begin{equation}
 \label{effectiveeddington}
 \Gammad={\sigmae\, \Ld \over 4\pi\, c\, G\,\Mwd},
 \end{equation}
 \noindent eq.~(\ref{masslossfromverticalwind1}) becomes
 \begin{equation}
 \label{masslossfromverticalwind2}
 \dot M=\simeq {\alpha\over 1-\alpha}\, \left({{3\sqrt{3}\over 2}
\,Q\,\Gammad}\right)^{1-\alpha\over\alpha} \; {\Ld \over c^2}.
 \end{equation}
 \noindent Up to the correction factors due to different gravity (and
geometry), this equation is identical to the CAK mass loss rate from a
point star, eq.~(\ref{gayleysmasslosseq}). Note that a disk wind is
more efficient in carrying mass loss than an O~star wind by a factor
$(3\sqrt{3}/2)^{(1-\alpha)/\alpha}$, which is due to the lower gravity
potential well. Re-writing eq.~(\ref{masslossfromverticalwind2}) as
$\dot M \equiv N\,\Ld/c^2$ (where the coefficient $N$ depends on
$\Ld$), and using the relevant parameters for novalike CVs introduced
above, gives $N\simeq 2$ and a disk mass loss rate $\dot M\simeq
10^{-12} \msy$, for $Q=2,000$.

\subsection{Tilted Disk Winds}
\label{sectionmasslosstiltedwind}

A more realistic picture of disk mass loss consists of a tilted wind
from a disk with radial temperature stratification. From eq.~(I.17),
using eq.~(\ref{effectiveeddington}),
 \begin{equation}
 \label{masslosstiltedwind}
 \dot M= {\alpha\over 1-\alpha}\, \left({{3\sqrt{3}\over 2}
\,Q\,\Gammad}\right)^{1-\alpha\over\alpha} \, D \,\langle{\dot
m}\rangle\, {\Ld \over c^2},
 \end{equation}
\noindent where
 \begin{equation}
 \langle{\dot m}\rangle \equiv \left. \int_{r_1}^{r_2} {dr\over r} \,
\dot m_r \, \bigl[r^2 \Flux_z(r,0) \bigr]^{1/\alpha} \right/
\int_{r_1}^{r_2} {dr\over r} \bigl[r^2 \Flux_z(r,0)\bigr]^{1/\alpha},
 \end{equation}
 \noindent and $\dot m_r$ is normalized in units of mass loss from a
vertical wind; $\langle{\dot m}\rangle$ is then a normalized mass loss
rate per $dr$, averaged over the wind base. Using values of $\dot m_r$
from Table~I.1, we estimate $\langle{\dot m}\rangle = 1.2$ for
$\alpha=2/3$, and $\langle{\dot m}\rangle = 2.2$ for $\alpha=1/2$,
both for SHS and Newtonian disks. For the latter, one has
$D=\ln(r_2/r_1)/ [\ln(\rd/\rwdt)]^{1/\alpha}$. This expression is
roughly correct also for the SHS disk, where $D$ cannot be calculated
analytically.  Inserting the values of $\langle{\dot m}\rangle$ and
$D$ into (\ref{masslosstiltedwind}) one finds, for $\alpha=1/2$ and
$2/3$, and for SHS and the Newtonian disks, assuming once again
typical parameters for novalike CVs and $Q=2,000$, that $\dot M \simeq
10^{-12} \msy$.

Interestingly, the mass loss rates for a vertical wind above an
isothermal disk and for tilted winds above disks with temperature
stratification are very similar. This means that the disk mass loss
rate is only a weak function of the tilt angle, {\it as long as} the
latter is aligned with the radiation flux. For the disk types used in
the present work, this range encompasses $\lambda\sim 50-90\arcdeg$,
according to Table~I.1.

It is readily shown that the mass loss rate for a LDW due to a single,
optically thick line is roughly $\Ld/c^2$. The above $N=2$, therefore,
implies that only few lines become optically thick in the present CV
disk wind model. Unlike for disk winds, in O~star winds of the order
of 100 lines become optically thick according to
eq.~(\ref{gayleysmasslosseq}).

Note that in Paper~I, eigenvalues $E$ were derived {\it without}
including the saturation of the force multiplier at $M_{\rm max}(t)=Q$
(i.e., without applying the exponential line-list cutoff of Owocki,
Castor, \& Rybicki 1988). The above mass loss rates are, therefore,
upper limits.

Finally, we derive the dependence of the mass loss rate $d\dot M$ on
$r_0$. According to (I.17), this relation is determined by the disk
temperature stratification and the run of $E\cri$ with $r_0$.  For an
isothermal disk, $E\cri$ was found to be independent of $r_0$, hence
$d\dot M/dr_0 \propto r_0^{(2-\alpha)/\alpha}$. Such an unrealistic
growth of $d\dot M/dr_0$ with radius is a consequence of the increase
of radiation energy with $r_0$. Alternatively, for the Newtonian disk,
fitting $E\cri(r_0)$ from Table~I.1 with a power law, one finds
approximately $d\dot M/dr_0 \propto r_0^{-0.8}$ for $r_0\gap
5\rwd$. The total disk mass loss rate scales therefore roughly as
$\dot M\propto \ln r_2$. For the SHS disk, approximately $d\dot
M(r_0)/dr_0 \propto r_0^{-1.9}$ for $r_0\gap 5\rwd$. At smaller $r_0$, the
radial dependency is weak for both types of disks. Therefore, the mass
loss rate $\dot M$ is centrally concentrated, and more so for SHS
disks.

\section{Velocity laws for disk winds}
\label{velocitylaws}

We discuss the wind velocity law by solving the Euler equation, first
neglecting, then accounting for ionization effects in the flow. Only
the flow above the SHS disk is analyzed in this section.

\subsection{Solutions to the algebraic Euler equation}

The geometrical expansion terms, discussed in Section 5.1 of Paper~I,
introduce a velocity dependency into the wind Euler equation.
However, we find that these terms leave the mass loss rates in the
model practically unchanged, and increase the wind velocity by at most
10\%. The reason for this is the small angle dispersion in the wind
streamlines. We, therefore, omit geometrical expansion terms here.

Without explicit dependence on velocity, the Euler equation becomes
purely algebraic. Figure~I.8 displays the solutions $W'(\XX)$ of the
Euler equation (I.20) for a tilted disk wind, at different $r_0$. The
velocity field above the disk is obtained by integrating $W'$, and is
displayed for a number of streamlines in Fig.~\ref{velocitylaw}, for
the SHS disk and $\alpha=2/3$. As shown in Paper~I, the LDW velocity
law is not a monotonic function of $\XX$. Clearly, the deceleration
regime (marked with a cross in Fig.~\ref{velocitylaw}) is unimportant
for streamlines starting at large $r_0$, because it lies at large
$\XX$ where the flow moves much faster than the local escape
speed. The inner streamlines do show a more pronounced kink.

The wind terminal velocities are found to be independent of $r_0$, for
$\alpha=1/2$, and linearly dependent on $r_0$ for $\alpha=2/3$,
growing by a factor of five over the whole disk. However, the above
examples may be of academic interest only, since the wind is expected
to go through an optically thick-to-thin transition beforehand. Hence
the actual observable terminal velocities may be smaller. Note that we
presume that the wind geometry stays unchanged at all $\XX$, and
neglect the saturation effect in the force multiplier at very small
optical depths in the wind. Both effects can influence the asymptotic
wind dynamics.

The velocity profiles in Fig.~\ref{velocitylaw} are well approximated
by the empirical velocity law used by Shlosman \& Vitello (1993) in
fitting the line profiles of novalike CVs,
 \begin{equation}
 \label{shlosmanvelocitylaw}
v=v_\infty \, {(X/X_{\rm acc})^\beta \over 1+(X/X_{\rm acc})^\beta}.
 \end{equation}
 \noindent Here, $X_{\rm acc}$ is the acceleration length scale along
the wind cone. We find the best fits to Fig.~\ref{velocitylaw} for
$\beta= 1.5-1.9$. Vitello \& Shlosman quote rather similar values of
$1.3-1.5$. Furthermore, $X_{\rm acc}\approx 1$ in
Fig.~\ref{velocitylaw}, which means that the footpoint radius $r_0$
sets the acceleration length. The reason for this is that $X_{\rm
acc}$ is determined by the effective gravity and disk radiation field,
i.e., by the auxiliary functions $g$ and $f$ (Paper~I), which change
on length scales $\sim r_0$. For this reason, we do not expect $X_{\rm
acc}$ to depend on $\alpha$. The $X_{\rm acc}$ from observed P~Cygni
line profiles are found to be somewhat larger, namely in the range
$1-10$ (Hoare \& Drew 1993; Vitello \& Shlosman 1993), depending on
individual objects.

\subsection{Effects of ionization stratification in the wind}
\label{abbottsdelta}

The major concern of the present model is the small optical depth in
the wind, where the latter appears to be semi-transparent even in the
strongest driving lines, when $Q=2,000$ is used. A possible resolution
of this problem may be related to our neglect of ionization structure
in the wind, which is expected to lead to a shallower velocity law and
smaller velocity gradients. To proceed, we parameterize the ionization
stratification in the simplest possible way, by introducing the
$\delta (> 0)$ parameter customarily used in stellar wind theory
(e.g., Abbott 1982), namely
 \begin{equation}
 \label{abbottdelta}
 g_{\rm L} \propto t^{-\alpha} \xi^{-\delta},
 \end{equation}
 \noindent where $\xi\equiv J/n_{\rm e}$ is the ionization parameter,
with $J$ being the frequency-integrated mean intensity.  Higher
ionization stages typically harbor fewer lines than lower ionization
stages, and therefore lead to a smaller line force.  Typical values
for O~stars are $\delta \lea 0.1$ (Abbott 1982; Puls 1987;
Pauldrach~et~al.~1994), but values as large as $\delta = 0.7$ have
been suggested recently for winds at low effective temperatures of
8,000~K (Kudritzki~et~al.~1998).

To include the $\delta$ correction term in the present model, we
calculate first $dJ$ from a disk annulus of width $dq$,
 \begin{equation}
 \label{meanintring} 
 dJ(r,z)=4\, I(q,0)\,q\,dq\,z\, {\sqrt{q^2+r^2+z^2-2rq}
\over (q^2+r^2+z^2)^2-(2rq)^2} \,\, E_2\left(-{4rq \over (q-r)^2+z^2}
\right),
 \end{equation}
 \noindent where $I$ is the isotropic intensity, and $E_2$ is the
complete elliptical integral of the second kind (Abramowitz \& Stegun
1965). For the temperature stratifications of interest, this
expression for $dJ$ cannot be integrated analytically over $q$ to give
$J$. Resorting to a 1-D numerical integration, Fig.~\ref{jandf} shows
$J$ and $\Flux$ as function of $\XX$ for the SHS disk at some
representative $r_0$. Importantly, for the outer wind, the critical
point lies upstream of the maximum of both $J$ and $\Flux$. Since $J$
increases along the streamline while the electron density drops, $\xi$
{\it increases} all the way to the critical point, as in O~star
winds. The line force at the critical point is, therefore, smaller for
$\delta>0$ than for $\delta=0$, and the same is true for the mass loss
rate. Alternatively, in the unlikely situation that the wind
recombines along the streamline, the driving force as well as the mass
loss would increase, because of a larger number of metal transitions.

In order to understand the effect of $\delta$ on the wind velocity
law, we include the $\delta$-correction (or ionization stratification
in the wind) above the critical point but neglect it below the latter,
thereby leaving the mass loss rate unaltered. Since the $\delta$-term
introduces a dependence of the line force on $W$, the solution to the
Euler equation has to be iterated until convergence is achieved.

We find that, assuming $\alpha=2/3$ and $\delta=0.2$, terminal speeds
decrease by a factor of $\sim 2$ for outer regions of the SHS disk,
whereas the optical depth $t$ increases by a factor of $\sim 4$,
thereby pushing the solution further away from the cutoff at $t\sim
10^{-7}$.  For inner disk regions, the effect of $\delta$ on
$v_\infty$ and $t$ is less pronounced. The somewhat ambivalent
conclusion is, therefore, that $\delta$-terms may, as desired, raise
$t$ by a factor of a few, but, at the same time also can lead to an
unwanted reduction in terminal speeds.

\section{Discussion}
\label{discussion}

In this section, we compare the mass loss rates from our model with
values quoted in the literature, which are typically estimated from
P~Cygni line fits or from dynamical wind modeling. We also mention
briefly some processes neglected in this work, which may have an
effect on $\dot M$.

\subsection{Comparison with $\dot M$ from kinematical wind modeling}

The most reliable estimates for mass loss rates from cataclysmic
variables are so far from P~Cygni line fits. For novalike variables
with parameters similar to those considered here, Vitello \& Shlosman
(1993) and Knigge~et~al.~(1995) find a lower limit of $\sim 10^{-2}$
to the ratio of mass loss-to-accretion rates. Below this value, the
observed line profiles cannot be reproduced because the wind becomes
over-ionized. This is about two orders of magnitude larger than $\dot
M$ derived in Section~\ref{diskwindmasslossrates}, when $Q=2,000$ is
used as in O~stars.  It is important, however, that in the line
fitting a highly idealized blackbody disk spectrum was used. This is a
clear overestimate of the ionizing radiation flux shortward of the
Lyman edge (Polidan~et~al.~1990; Long~et~al.~1991, 1994; van
Teeseling~et~al.~1993; Knigge~et~al.~1997). For a more realistic EUV
radiation field which accounts for the Lyman continuum cutoff, the
wind mass loss rate can be reduced, while maintaining the same
ionization parameter. We have been able to produce acceptable P~Cygni
line profiles down to $\dot M\sim 10^{-11}\msy$, for the present
system parameters. Essentially all the carbon was then in the form of
C~{\sc iv}, due to Auger ionization of C~{\sc ii} by X-rays.

\subsection{Comparison with $\dot M$ from dynamical wind modeling}

Next, we compare the mass loss rates from our analytical wind model
with those from dynamical simulations of disk winds.  The only
realistic dynamical modeling of CV disk winds attempted so far was
performed by PSD. We have provided a general comparison between our 
models in Paper~I, and turn here to mass loss rates. 

Using $k=0.2$ and $\alpha=0.6$, PSD find for the SHS disk mass loss
rates of $\dot M=5\times 10^{-14} \msy$ and $\dot M=5\times 10^{-12}
\msy$, corresponding to disk luminosities of $L=8\, \Lsun$ and $L=24
\,\Lsun$, respectively. It is clear that such an increase by a factor
of 100 in $\dot M$ cannot be understood from the simple CAK scaling,
where $\dot M\propto L^{1/\alpha}$. Instead, this strong dependence on
$\Ld$ is a consequence of the optically thick-to-thin transition in
the disk LDW when $Q\,\Gamma \sim 1$. The reason for this is that PSD
apply the exponential line-list cutoff (Owocki~et~al.~1988), due to
which the force multiplier $M(t)$ reaches a maximum value in optically
thin flow regions. This suppresses the mass loss rate as compared to
the case of a pure power law force multiplier.

Even, for a low-luminosity disk, PSD find that large $\dot M$ can be
driven when $\alpha$ is as large as 0.8. However, PSD assume the same
value of $k=0.2$ for {\it all} $\alpha$. As is evident from Fig.~4 of
Gayley (1995), $k$ drops by a factor of 10 when $\alpha$ increases
from 0.6 to 0.8, leading to a very similar $\dot M$ in both
cases. This is a consequence of $Q\, \Gamma \sim 1$ for a disk wind in
eq.~(\ref{masslosstiltedwind}). Contrary to this, for O~star winds, $Q
\Gamma \gg 1$, hence $\dot M$ depends strongly on $\alpha$.

For a more luminous disk with $\Ld=24 \, \Lsun$, we estimate $\dot
M\simeq 4\times 10^{-12} \msy$ from eq.~(\ref{masslosstiltedwind})
when a value of $D$ is used characteristic of the narrow wind base of
PSD. The agreement with $\dot M= 5\times 10^{-12} \msy$ as found by
PSD is very good.

\subsection{The mass loss paradigm in disk winds from CVs}
\label{neglectedprocesses}

Taken at face value, the modified CAK theory of LDWs from CV disks
predicts surprisingly low mass loss rates when a line force parameter
$Q=2,000$ is used as in O~stars. The calculated rates of $\sim
10^{-12} \msy$ for $\Ld=10\Lsun$ (or even lower, when the saturation
of the force multiplier is accounted for), imply that LDWs will have a
thick-to-thin transition in the strongest driving lines, and will have
difficulties in reproducing the observed line profiles. What are the
possible solutions to this problem?

Firstly, higher values of $Q$ are the most obvious way to increase
$\dot M$ in the wind. The value of $Q$ is highly
uncertain. Especially, it is unclear how the different spectral shape
of disk radiation and its effect on the non-LTE electron occupation
numbers in ions will modify $Q$ as compared to its O~star value. Such
calculations have never been attempted and are clearly beyond the
scope of this paper. Theoretically, the value of $Q$ can be significantly
larger than for the O~stars.

Analysis of O~star LDWs revealed $Q\,Gamma$ to be the key parameter
determining the mass loss rate in the wind (Gayley 1995). Namely, for
$Q\,\Gamma < 1$, the wind ceases to exist. However, for O~stars
$Q\,\Gamma > 100$, producing a kind of a `safety belt'. The situation
is different for winds from luminous CV disks, where $Q\,\Gamma\sim
1$, if the O~star value for $Q$ is used (this work and PSD). This was
already foreseen by Vitello \& Shlosman (1988), which claimed that the
disk winds appear to be much more restrictive than O~star winds.

Secondly, the CAK line force does not account for the ionization
stratification in the wind. So a situation can arise where the wind is
driven by the part of the disk which is UV bright, while the
ionization is controlled by the central disk region, e.g., the
boundary layer, which is X-ray bright. This will be especially
pronounced in the presence of a Lyman continuum cutoff in the UV
source. Under these circumstances, due to the wind over-ionization,
the line force will be reduced below its CAK value (MacGregor,
Hartmann, \& Raymond 1979; Fransson \& Fabian 1980) and hence the
velocity gradient as well. A shallow velocity law will increase the
overall optical depth in the wind and push the thick-to-thin
transition downstream, away from the critical point. It is important
in this context that the minimal changes in the P~Cygni line profiles
in eclipsing novalike CVs {\it require} very shallow velocity profiles
in the wind (e.g., Shlosman, Vitello, \& Mauche 1996).

Additional processes can affect the mass loss rates, however to a
lesser degree. Firstly, the assumption that each wind parcel is confined
to a straight cone is a major limitation of the present model. For
bended streamlines, a better alignment with the radiative flux vector
seems possible, which should increase the mass loss. Secondly,
Pauldrach \& Puls (1990) find a sudden increase in stellar mass loss
rates of B supergiants when the Lyman continuum becomes optically
thick at the wind base, e.g., when the effective temperature drops
below a certain threshold or if $\Gamma$ reaches some critical
value. This induces a shift to lower wind ionization, which in turn
increases the mass loss (more driving lines), increases the Lyman jump
even more and so on, until a stable situation with high $\dot M$ and
low wind ionization is reached. Thirdly, $Q$ depends linearly on the
wind metallicity, and enhanced metal abundances could lead to a larger
$Q\,\Gamma$, which governs the mass loss. This is significant in
particular to the He abundance due to the efficient convection in the
low-mass secondary stars, but may be relevant to metals as well.

\section{Summary}

The focus of this second paper on line-driven winds from accretion
disks in cataclysmic variables is on the theoretical estimates of the
mass loss rates and wind velocity laws. Our results are as follows.

The derived mass loss rate using the modified CAK formalism appears to
be substantially smaller than that inferred from P~Cygni line fits,
even with supressed Lyman continuum, and more so when the saturation
effect in the line force multiplier is included. Yet this rate is in a
good agreement with results from time-dependent, 2-D dynamical wind
simulations by PSD. The reason for low mass loss rates is that the key
parameter controlling LDWs, $Q\,\Gamma\sim 100$ for O~star winds, is
only $\sim 1$ for disk winds, when O~star values for $Q$ are
used. Some potential resolutions to this problem were proposed.

We find that the mass loss is dominated by the inner decade in disk
radii. For the Shakura \& Sunyaev and for Newtonian disks, the mass
loss per unit radius is roughly uniform out to five white dwarf radii,
and drops $\propto r^{-1.9}$ and $r^{-0.8}$ at larger $r$,
respectively.

Due to their low mass loss rates, CV winds should experience a
thick-to-thin transition even in the strongest lines. CV winds should,
therefore, resemble more closely winds of B~stars near-the-main
sequence than of O~supergiants.

The wind velocity profiles show a slowly accelerating flow, with a
characteristic acceleration length given by the footpoint radius of
the streamline in the disk. Fitting the observed line profiles using
kinematical models suggest even slower accelerating winds. The
observable terminal velocity of the wind is associated with the
thick-to-thin transition in the driving lines. Due to this latter
fact, and due to uncertainties in the ionization stratification and in
the anticipated streamline bending at large radii, the actual wind
terminal velocity is poorly constrained in our model.

\acknowledgments
We are grateful to Jon Bjorkman, Rolf Kudritzki, Chris Mauche, Norman 
Murray, Stan Owocki and Joachim Puls for numerous discussions on 
various aspects of line-driven winds. I.S. acknowledges hospitality
of the IGPP/LLNL and its Director Charles Alcock, where this work
was initiated. This work was supported in part by NASA grants NAG5-3841
and WKU-522762-98-06, and HST AR-07982.01-96A (I.S.), and was
performed under the auspices of the US Department of Energy by LLNL user
contract number W-7405-ENG-48 (P.V.).

\appendix

\section{Disk Eddington factors required for disk line-driven winds}

Using the eigenvalues $E\cri$ from Paper~I, one can further quantify
the disk Eddington factors required to launch a line-driven wind which
is optically thick at least up to the critical point. Consider first a
vertical wind above an infinite, isothermal disk. The Euler equation
(\ref{euler_justsowind}) becomes, for $\FF_x=1$ (the flux component
along ${\bf\hat\xx}$, normalized to the footpoint flux),
 \begin{equation}
 W'=-g +\Gamma(r_0) M(t_x).
 \end{equation}
 \noindent From eq.~(I.18) at the critical point,
 \begin{equation}
 \Gamma(r_0) = {2/3\sqrt{3} \over (1-\alpha) M\cri(t)}.
 \end{equation}
 \noindent Since the maximum of $M(t)$ for a gas of solar composition
is $\sim 2,000$ (Abbott 1982), one has that $\Gamma(r_0) \gap 4\times
10^{-4}$ for all $r_0$. Next, for the more realistic case of a tilted
wind above a disk with radial temperature stratification, the
simplified Euler equation~(\ref{euler_justsowind}) reads
 \begin{equation} 
 W'=-g+\Gamma(r_0)\, \FF_\xx\, M(t_\xx).
 \end{equation}
 \noindent Using $W'\cri=(\alpha/1-\alpha) \, g\cri$ and assuming
$g\cri\simeq \XX\cri$ for critical points close to the disk,
 \begin{equation}
 \label{forcemultipliergeneral}
 \Gamma(r_0) = {\XX\cri \over (1-\alpha) \, \FF_\xx(r_0,\XX\cri) \,
M\cri(t)}.
 \end{equation}
 \noindent According to Fig.~I.5, $\FF < 4$, hence $\Gamma(r_0)\gap
10^{-4}$ is required for the wind to be optically thick at the
critical point.

\clearpage

 \begin{figure}
 \epsscale{1.}
 \plotone{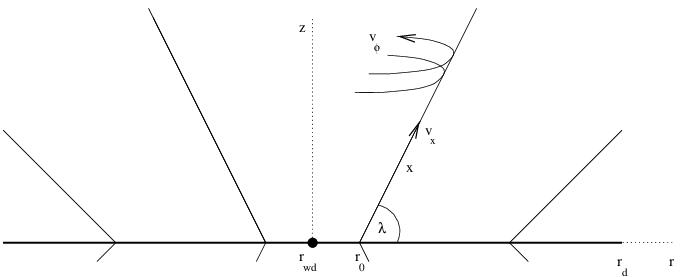} 
 \caption{\label{windcoordinates} Adopted flow geometry for a CV disk
wind. The streamlines are helical lines, and are assumed to lie on
straight cones.}
 \end{figure}

 \begin{figure}
 \epsscale{1.}
 \plotone{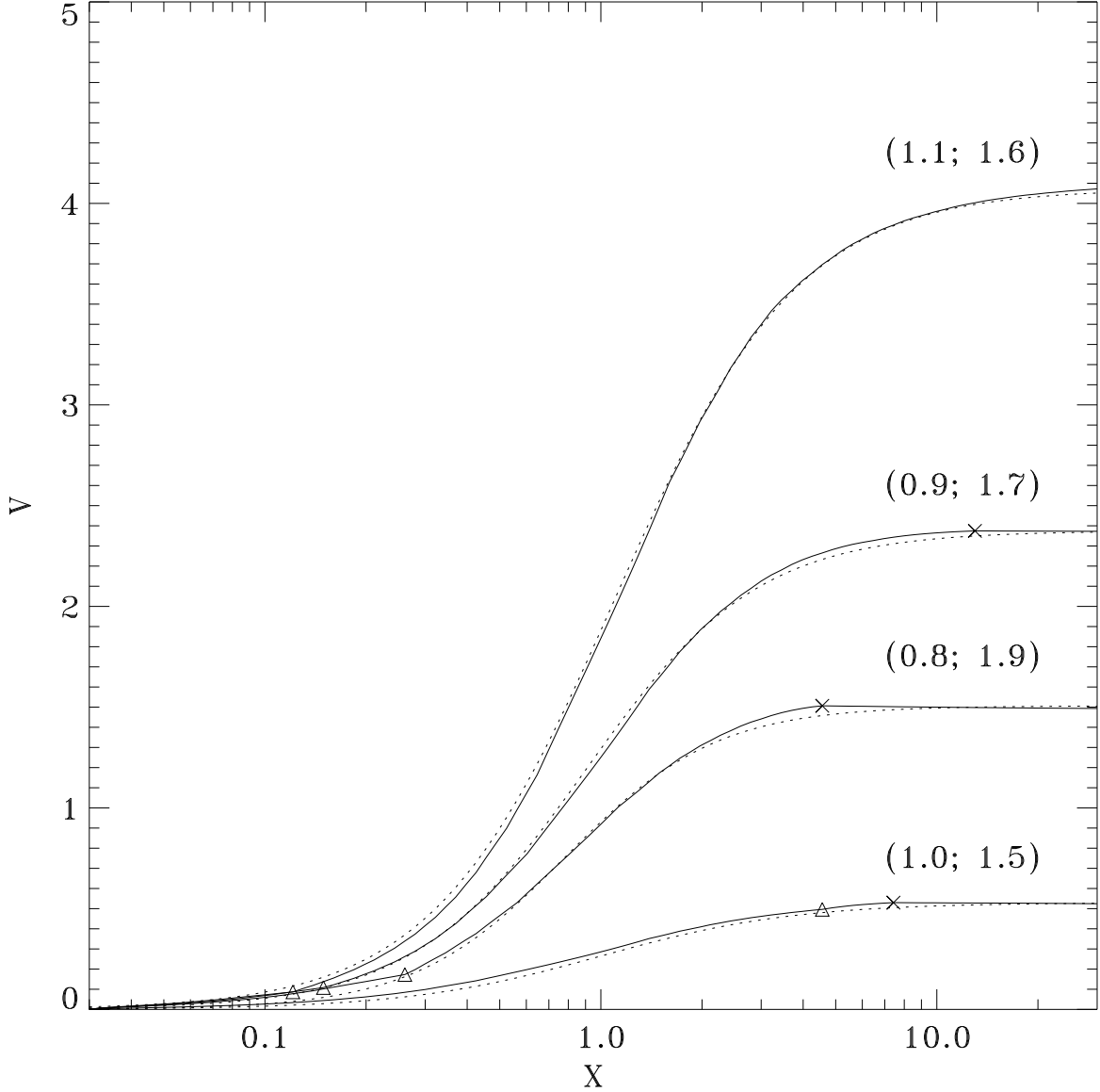}
 \caption{\label{velocitylaw} Velocity profiles for the SHS disk with
$\rd=30\rwd$, for $\alpha=2/3$, and at footpoint radii 3, 7, 10,
$15\rwd$ (from bottom to top). The dotted lines are fits using the
velocity law eq.~(\ref{shlosmanvelocitylaw}), with the values of
$(\XX_{\rm acc}; \beta)$ given in brackets. The velocity $V$ is in
units of the local escape speed at $r_0$. For the tilt angles,
eigenvalues $\lambda\cri$ from Paper~I were used. Triangles mark the
critical point, crosses the point where the wind starts to
decelerate.}
 \end{figure}

 \begin{figure}
 \epsscale{1.}
 \plotone{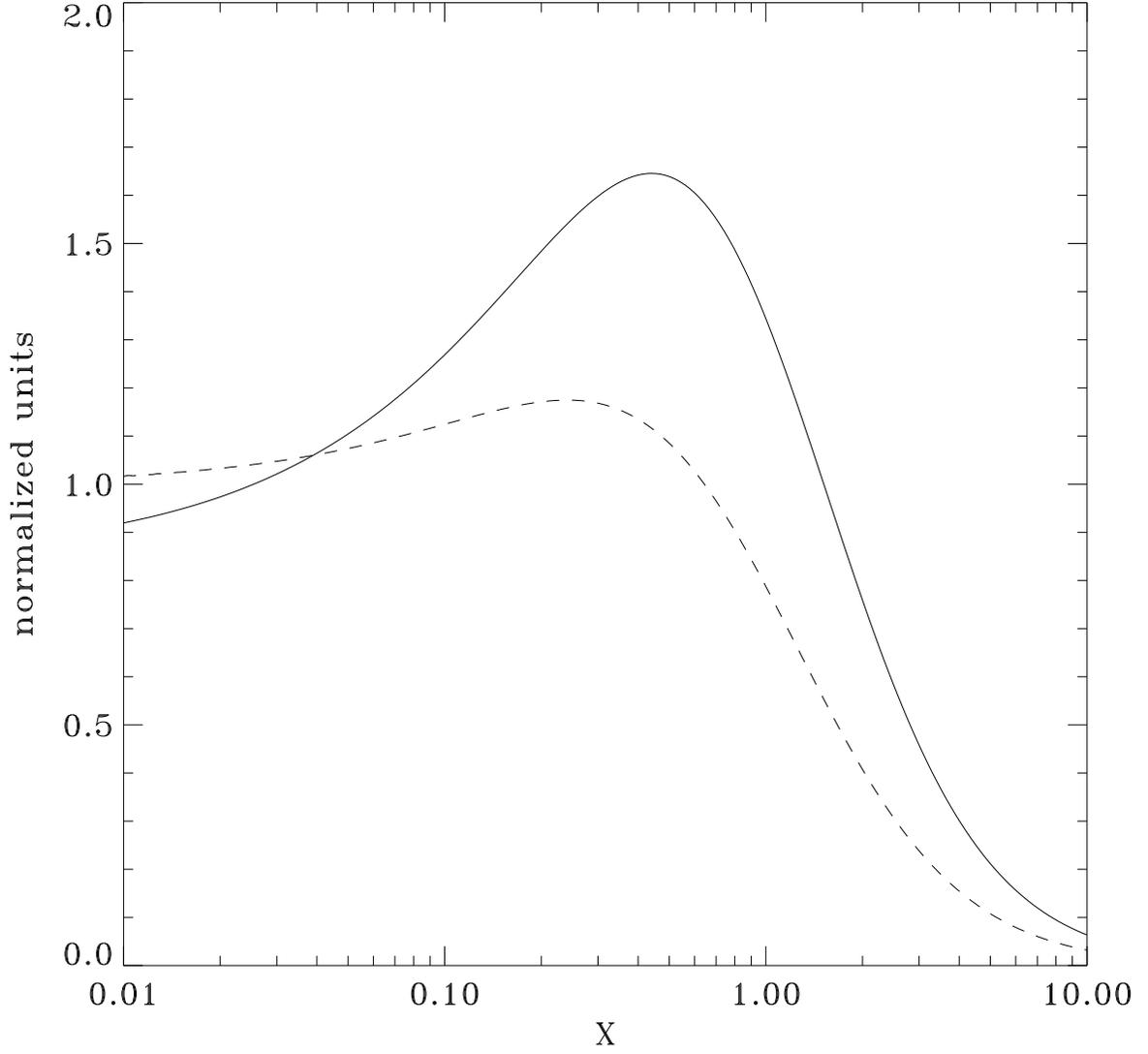}
 \caption{\label{jandf} Normalized, projected flux (full line) and
mean intensity (dashed line) above the SHS disk ($\rd=30\rwd$), along
a ray with footpoint $r_0=10\rwd$ and $\lambda\cri=58\arcdeg$.}
 \end{figure}

\end{document}